\documentclass[conference]{IEEEtran}
\IEEEoverridecommandlockouts
\usepackage{cite}
\usepackage[numbers,sort&compress]{natbib}
\usepackage{amsmath,amssymb,amsfonts,gensymb}
\usepackage{algorithmic}
\usepackage{graphicx}
\usepackage{textcomp}
\usepackage{booktabs}
\usepackage{multirow}
\usepackage{xcolor}
\def\BibTeX{{\rm B\kern-.05em{\sc i\kern-.025em b}\kern-.08em
    T\kern-.1667em\lower.7ex\hbox{E}\kern-.125emX}}

\begin{document}

\title{Sound Event Detection and Localization with Distance Estimation \\
\thanks{The authors wish to thank CSC-IT Centre of Science Ltd., Finland,  for providing computational resources.}
}

\author{\IEEEauthorblockN{Daniel Aleksander Krause, Archontis Politis, Annamaria Mesaros}
\IEEEauthorblockA{\textit{Faculty of Information Technology and Communication Sciences} \\
\textit{Tampere University}\\
Tampere, Finland \\
daniel.krause@tuni.fi, archontis.politis@tuni.fi, annamaria.mesaros@tuni.fi}
}

\maketitle

\begin{abstract}
Sound Event Detection and Localization (SELD) is a combined task of identifying sound events and their corresponding direction-of-arrival (DOA). While this task has numerous applications and has been extensively researched in recent years, it fails to provide full information about the sound source position. In this paper, we overcome this problem by extending the task to Sound Event Detection, Localization with Distance Estimation (3D SELD). We study two ways of integrating distance estimation within the SELD core - a multi-task approach, in which the problem is tackled by a separate model output, and a single-task approach obtained by extending the multi-ACCDOA method to include distance information. We investigate both methods for the Ambisonic and binaural versions of STARSS23: Sony-TAU Realistic Spatial Soundscapes 2023. Moreover, our study involves experiments on the loss function related to the distance estimation part. Our results show that it is possible to perform 3D SELD without any degradation of performance in sound event detection and DOA estimation.

\end{abstract}

\begin{IEEEkeywords}
Sound event detection, sound source localization, sound distance estimation, Ambisonics, binaural recordings
\end{IEEEkeywords}

\section{Introduction}
Computational Auditory Scene Analysis (CASA) has emerged as a prominent area of study in recent years \cite{Virtanen2018Ananlysis}. The automated examination of audio content holds substantial potential for diverse practical applications, including speech recognition \cite{speechrec}, autonomous robots \cite{4058525}, surveillance systems \cite{opatka2011ApplicationOV}, and support systems for the hearing-impaired \cite{5202720}. CASA encompasses a spectrum of audio-related tasks, ranging from acoustic scene classification \cite{Mesaros2018a} and audio tagging \cite{Fonseca2018_DCASE} to sound source localization \cite{krauseconv} and sound event detection \cite{Mesaros2019_TASLP}. Most tasks are currently approached by Deep Neural Network (DNN) models.

Although existing research has predominantly focused on individual tasks, an evolutionary progression towards the development of complex scene analysis systems involves the creation of models capable of simultaneously addressing multiple objectives. This progressive approach has been observed in recent research, as exemplified by the exploration of joint sound event detection and localization (SELD) \cite{Adavanne_2019}. SELD combines the task of identifying the temporal activities of sound events, altogether with their direction of arrival (DOA) and textual label. However, this approach does not take advantage of full spatial information by limiting it to the DOA only. In many cases, performing Sound Distance Estimation (SDE) would be also important to obtain the explicit position of the sound source in space. 

Research on DNN-based SELD has shown multiple ways of solving the problem of matching two tasks for different scenarios. In \cite{Adavanne_2019}, the authors proposed a Convolutional Recurrent Neural Network (CRNN) with a two branch solution, in which SED and DOA estimation are solved with independent classwise outputs. To allow for detection of multiple events at the same time, a track-wise output has been proposed in \cite{9413473, 9053045}. The activity-coupled Cartesian DOA (ACCDOA) scales the Cartesian DOA vector by its corresponding event activity, overcoming the need for a multi-task approach \cite{shimada2021accdoa}. Finally, the multi-ACCDOA method takes advantage of the trackwise approach and the ACCDOA format, allowing for independent detection of same-class events \cite{9746384}. 

Regarding SDE, studies on DNN-based approaches have been limited mostly to the binaural format. Most of them use a classification approach, in which the distance is expressed as a finite set of pre-defined distances in the close area up to 4 meters \cite{yiwere2019, sobhdel2021}. In \cite{kushwaha2023sound}, the authors studied multiple loss functions to perform distance estimation with an activity detection branch for a tetrahedral microphone array. Few studies investigated the performance of distance estimation in conjunction with DOA estimation \cite{Yiwere2017DistanceEA, krause2021joint, Krause2024Moving}, however no approach has been made to merge SDE with SELD.

In this paper, we investigate the joint task of Sound Event Detection, Localization and Distance Estimation. We study two ways of performing all three tasks jointly. First, we examine a multi-task approach, in which two separate branches are responsible for SELD and SDE. Second, we propose the multi activity-coupled Cartesian Distance and DOA (\textbf{multi-ACCDDOA}) method, which is an an extension of the known multi-ACCDOA format, by including the distance in the estimated vector. For both approaches, we study the influence of several loss functions to determine which is the most suitable for the joint task. Experiments are conducted for both First Order Ambisonics (FOA) and binaural recordings to investigate the potential performance of the task in a more limited audio format. To the authors' knowledge, this is the first study investigating joint modelling of all three tasks.

\section{Method}
\label{method}
\subsection{Features}
\label{features}

\begin{table}[b] 
\vspace*{-5mm}
\caption{Input parameters for the models.}
\vspace*{-4mm}
\begin{center}
\begin{tabular}{l c c c c}
\toprule
\textbf{Audio data format} & \textbf{CH} & \textbf{T} & \textbf{F} & \textbf{P} \\
\midrule
Ambisonics & 7  & 250  & 64 & [4, 4, 2] \\
Binaural & 4  & 250  & 512 & [8, 8, 4] \\
\bottomrule
\end{tabular}
\label{tab:model_inputs}
\end{center}
\vspace*{-5mm}
\end{table}

A feature input matrix of shape $CH \times T \times F$ is fed to the model, where $CH$, $T$ and $F$ stand for the number of channels, time sequence length in frames and number of features respectively. The set of features utilized to train the models depends on the audio format under investigation as summarized in Table \ref{tab:model_inputs}. In this paper, we study the Ambisonic and binaural formats. Each file is split into clips of length $T=250$. A complex spectrogram of the signal is obtained using a Short-Time Fourier Transform (STFT) with a Hamming window of length 40~ms and 50\% overlap. This results in $F=512$ frequency bins. 

For Ambisonics, we accumulate the magnitude spectrograms from 4 channels into $F=64$ mel energies. In order to explore spatial cues, we extract 3 intensity vector matrices as in \cite{yasuda2020sound}, accumulated along the mel frequencies to fit the required number of features. The overall input matrix sums up to $CH=7$ feature channels. 

For the binaural format, we extract the mean magnitude spectrogram from both binaural channels. To represent spatial information about the signal, we extract sines and cosines of Interaural Phase Differences (IPD), which provide a smooth representation of phase values and avoid phase wrapping. This feature has been shown to perform successfully in binaural DOA estimation and SDE \cite{Krause2024Moving, garcia2022binaural}. On top of that, we use ILDs, which constitute another major binaural cue that becomes important above 1.5~kHz. This set of features results in $CH=4$ feature channels that are fed to the model.

\subsection{Model}
\label{model}

\begin{figure}[!t]
    \centering
        \includegraphics[width=\linewidth]{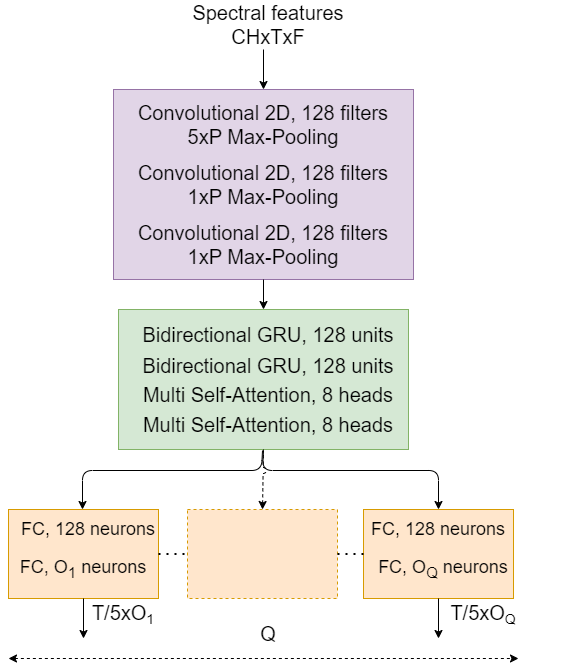}
        \vspace*{-5mm}
    \caption{Architecture of the deep neural network. }
    \label{fig:architecture}
    \vspace*{-5mm}
\end{figure}

We employ a convolutional recurrent neural network (CRNN) model type which is common for SELD. To perform 3D SELD, we modify the model outputs to contain the included distance estimation part. The architecture of the utilized model is depicted in Fig. \ref{fig:architecture}. First, the feature input matrix is processed by three 2D convolutional blocks, each consisting of 128 filter kernels, batch normalization and max-pooling across the feature dimension. Additionally, the first layer involves pooling across the time dimension with the rate of 5. The pooling rate across the featue dimension depends on the utilized feature set and is meant to reduce the feature dimension to 4 in the last convolutional block. Hence, for Ambisonics the rates are equal to $P=[4,4,2]$, whereas for the binaural format it is $P=[8,8,4]$. Next, the feature maps are passed to two bi-directional gated recurrent units (GRUs) and two Multi-Head Attention layers, with 8 heads each.

The output of the model consists of $Q$ branches, where $Q$ depends on the number of performed tasks. Each branch is composed of two fully connected (FC) layers, the former containing 128 neurons, and the latter outputting $C_q$ values. The parameters of the last layers are determined by the utilized method of including SDE within the 3D SELD framework. Here, we propose two techniques:

\textbf{I. Multi-ACCDDOA:} we modify the single task multi-ACCDOA approach proposed in \cite{9746384}. Compared with the former, we extend the 3-element DOA vector to include the distance estimate as well. For $N$ tracks, $C$ classes and $T$ frames, we define the output as \begin{math}y_{nct}=[a_{nct}R_{nct}, D_{nct}] \end{math}, where $n, c, t$ indicate the output track number, target class and time frame,  $a_{nct}\in\{0, 1\}$ stands for the detection activity, $R_{nct}\in\langle-1,1\rangle$ are to the DOA vectors and $D_{nct}\in\langle0,\infty)$ corresponds to distance values. The dimensions hold the following characteristics: \begin{math}\textbf{a, D} \in \mathbb{R}^{NxCxT}, \textbf{R} \in \mathbb{R}^{3xNxCxT}\end{math}, and $||\textbf{R}_{nct}||=1$. We model up to $N=3$, hence $O_{1}=156$. The whole output is linear to contain the range of both DOA and distance values. The multi-ACCDDOA model is trained using Auxiliary Duplicating Permutation Invariant Training (ADPIT) as in \cite{9746384}. The final loss function is defined as:
\begin{equation}
    \mathcal{L}^{ADPIT}=\frac{1}{CT}\sum_{c}^{C}\sum_{t}^{T}\min_{\alpha\in Perm[ct]}l_{\alpha,ct}^{ACCDDOA},
\end{equation}
\begin{equation}
    l_{\alpha,ct}^{ACCDDOA}=\frac{1}{N}\sum_{n}^{N}\mathcal{L}(y_{\alpha,nct},\hat{y}_{\alpha,nct}),
\end{equation}
where $\mathcal{L}(\cdot)$ is a loss function of choice, $\alpha$ is one possible track permutation and $Perm[ct]$ is the set of all possible permutations.

\textbf{II. Multi-task (MT):} here, the output is split into $Q=2$ branches. The first branch performs SELD using the classwise ACCDOA approach as described in \cite{shimada2021accdoa}. The output is defined as \begin{math}y_{1,ct}=a_{ct}R_{ct}\end{math}. Since the models estimate the DOA vector as $x,y,z$ Cartesian coordinates for each class and $C=13$, this results in $O_{1}=39$. The second branch is responsible for classwise SDE, where  \begin{math}y_{2,ct}=D_{ct}\end{math} and $O_{2}=13$. The ACCDOA output is normalized with a tanh activation, whereas the distance branch uses a Rectified Linear Unit. The matrix dimensions are defined as follows: \begin{math}\textbf{a, D} \in \mathbb{R}^{CxT}, \textbf{R} \in \mathbb{R}^{3xCxT}\end{math}. The final loss is a sum of both branches:
\begin{equation}
    \mathcal{L}^{MT}=\frac{1}{CT}\sum_{c}^{C}\sum_{t}^{T}(\mathcal{L}_{1}(y_{1,ct},\hat{y}_{1,ct})+\mathcal{L}_{2}(y_{2,ct},\hat{y}_{2,ct})),
\end{equation}
where $\mathcal{L}_{1}(\cdot)$ and $\mathcal{L}_{2}(\cdot)$ are the losses of the both branches.

We note that the output of the MT approach does not allow for differentiating between overlapping sources of the same class, whereas the multi-ACCDDOA approach overcomes this problem. The output parameters are summarized in Table \ref{tab:model_outputs}. Models are implemented in PyTorch \cite{paszke2019pytorch} and trained using the Adam optimizer for 250 epochs with 75 epochs of patience. 

\begin{table}
\caption{Output parameters for different models.}
\vspace*{-4mm}
\begin{center}
\begin{tabular}{l c c c}
\toprule
\textbf{Method} & \textbf{Q} & \textbf{O\lowercase{q}} & \textbf{Output activation} \\
\midrule
\midrule
\hspace{1mm}Multi-task & 2 & [39, 13] & [tanh, ReLU] \\
\hspace{1mm}Multi-ACCDDOA & 1 & 156 & linear \\
\bottomrule
\end{tabular}
\label{tab:model_outputs}
\end{center}
\vspace*{-5mm}
\end{table}

\subsection{Loss functions}
\label{lossfunctions}
In recent works on SELD using ACCDOA, the mean squared error (MSE) has been established as the most common loss function. However, the inclusion of the distance estimation part introduces a different value range, for which other loss functions might be more appropriate. The standard MSE function prioritizes sound sources which are further away from the origin, since large distances create a more significant error. In \cite{kushwaha2023sound} the authors introduced a relative regressor function, which evens out the error across ground truth distances, therefore penalizing the SDE branch more fairly. Here, we investigate the proposed loss functions for a joint 3D SELD model. The investigated functions include:

\begin{itemize}
    \item \textbf{Mean Squared Error:} 
    
    \begin{math}
    MSE = \frac{1}{M}\sum_{m=0}^{M-1}(y[m]-\hat{y}[m])^2,
    \end{math}
    \item \textbf{Mean Absolute Error:} 
    
    \begin{math}
    MAE = \frac{1}{M}\sum_{m=0}^{M-1}|y[m]-\hat{y}[m]|,
    \end{math}
    \item \textbf{Mean Square Percent Error:} 
    
    \begin{math}
    MSPE = \frac{1}{M}\sum_{m=0}^{M-1}(\frac{y[m]-\hat{y}[m]}{\hat{y}[m]})^2,
    \end{math}
    \item \textbf{Mean Absolute Percent Error:} 
    
    \begin{math}
    MAPE = \frac{1}{M}\sum_{m=0}^{M-1}|\frac{y[m]-\hat{y}[m]}{\hat{y}[m]}|,
    \end{math}
\end{itemize}
where $M$ stands for the number of estimated values, $y$ and $\hat{y}$ correspond to the predicted and ground truth values. 
For the MT model, we investigate all aforementioned loss functions for the SDE branch ($\mathcal{L}_{2}$). Since scaling the DOA vectors with distance values would put a larger weight to close sources, the ACCDOA branch ($\mathcal{L}_{1}$) keeps the MSE loss function for all scenarios. For similar reasons, the multi-ACCDDOA approach ($\mathcal{L}$) is tested only for the MSE and MAE losses.

\section{Experiments}
\label{experiments}
\subsection{Data}
\label{data-part}

For experiments, we use the Ambisonic audio-only version of the STARSS23 dataset. The dataset includes 7 hours and 22 minutes of real recordings, which is split intro training data (90 clips) and testing data (78 clips). There are 13 sound event classes present, including female speech, male speech, clapping, telephone, laughter, domestic sounds, footsteps, door, music, musical instrument, water tap, bell and knock. The scenarios include up to 3 overlapping sound sources. For more details, refer to the original paper \cite{Shimada2023starss23_arxiv}. In order to increase the amount of training data, we synthesized more mixtures using the data generator described in \cite{politis2021dataset}. The amount of additional data sums to 1200 clips of one-minute mixtures with the same maximum polyphony level as the original dataset. The data was synthesized using sounds from FSD50k \cite{fonseca2022fsd50k}. In order to provide experiments for the binaural format, we convert the whole dataset from Ambisonic to binaural using the spaudiopy library \cite{spaudiopy}. Decoding is performed via magnitude least squared matching to a measured set of Head Related Transfer Functions as described in \cite{ambisonics}.

\subsection{Evaluation metrics}

To evaluate our models, we use the SELD metrics from the DCASE Challenge 2023 Task 3. These include the Error Rate (ER), $F_{1}$ score for SED, and the DOA error and the localization recall for localization. The detection metrics are location aware, i.e., the events are counted only if the assigned source falls within ±20° of the ground truth DOA, whereas localization metrics are calculated only for true positives. On top of the well established SELD metrics, we add the distance error, which is defined as the mean absolute error between ground truth and predicted distances. The metrics are calculated in one second segments using micro-averaging and the matching between ground truth and predictions is done via the Hungarian algorithm referring to the angular distance between sources. For more details, see \cite{politis2020overview}.

\subsection{Results}
\label{results}

\begin{table*}   
  \centering
  \caption{Results obtained for Ambisonics.}
  \vspace*{-2mm}
    \begin{tabular}{c|c|c||c|c||c|c||c}
    \toprule
    Method & SELD loss & Dist. loss & ER & $F_{1} [\%]$ & DOA error [°] & Recall $[\%]$ & Dist. error [m] \\
    \midrule
    \midrule
    \multirow{4}{*}{Multi-task} & MSE & MSE & \textbf{0.63 [0.59, 0.67]}	& 41.4 [36.58, 47.04] &	22.5 [19.18, 25.58] & 61.0 [57.46, 65.57]	& 0.95 [0.82, 1.05] \\
    & MSE & MAE & 0.64 [0.60, 0.68]	& 43.6 [39.18, 48.60] & \textbf{21.6 [18.63, 24.36]} & 41.10 [37.36, 45.64] & 0.93 [0.80 , 1.02] \\
    & MSE & MSPE & 0.63 [0.59, 0.68] & \textbf{44.1 [38.92, 48.72]} & 23.2 [18.31, 28.13] & \textbf{64.7 [61.64, 68.68]} & 0.89 [0.77 , 0.99] \\
    & MSE & MAPE & 0.65 [0.61, 0.68] & 43.5 [38.97, 47.80] & 22.0 [18.94 , 24.80] & 64.5 [61.18, 68.72] & \textbf{0.88 [0.75, 0.97]} \\
    \midrule
    \multirow{2}{*}{Multi-ACCDDOA} & \multicolumn{2}{c||}{MSE} & \textbf{0.65 [0.61, 0.70]}	& \textbf{44.2 [39.45, 48.65]} & 22.9 [19.33 , 26.46] & \textbf{68.4 [65.15, 72.33]} & 0.92 [0.80 , 1.01] \\
    & \multicolumn{2}{c||}{MAE} & 0.86 [0.82, 0.91] & 21.5 [13.98, 28.47] & \textbf{17.7 [14.09 , 21.05]} & 19.1 [12.44, 24.90] & \textbf{0.74 [0.54 , 0.93]} \\
    \bottomrule
    \end{tabular}
  \label{tab:ambi_results}
  \vspace*{-4mm}
\end{table*}
\begin{table*}   
  \centering
  \caption{Results obtained for binaural audio.}
  \vspace*{-2mm}
    \begin{tabular}{c|c|c||c|c||c|c||c}
    \toprule
    Method & SELD loss & Dist. loss & ER & $F_{1} [\%]$ & DOA error [°] & Recall $[\%]$ & Dist. error [m] \\
    \midrule
    \midrule
    \multirow{4}{*}{Multi-task} & MSE & MSE & \textbf{0.82 [0.79, 0.86]} & \textbf{20.0 [15.40, 24.40]} & 41.1 [34.63 , 47.77] & \textbf{45.6 [41.89, 49.02]} & 1.02 [0.90 , 1.12] \\
    & MSE & MAE & 0.85 [0.81, 0.87]	& 16.5 [13.91, 20.29] & 38.6 [32.40 , 43.28] & 36.7 [33.61, 40.89] & 1.04 [0.90 , 1.15] \\
    & MSE & MSPE & 0.85 [0.81, 0.89] & 19.3 [15.70, 23.91] & 38.9 [31.91 , 44.28] & 38.9 [35.52, 43.56] & 1.01 [0.87 , 1.12] \\
    & MSE & MAPE & 0.87 [0.84, 0.91] & 18.5 [15.16, 22.23] & \textbf{38.1 [32.77 , 42.45]} & 42.2 [38.59, 45.83] & \textbf{0.98 [0.86 , 1.09]} \\
    \midrule
    \multirow{2}{*}{Multi-ACCDDOA} & \multicolumn{2}{c||}{MSE} & \textbf{0.87 [0.82, 0.91]} & \textbf{21.1 [17.38, 25.26]} & \textbf{39.7 [32.58 , 46.34]} & \textbf{48.0 [44.75, 51.25]} & 0.99 [0.88 , 1.09] \\
    & \multicolumn{2}{c||}{MAE} & 0.97 [0.94, 0.99] & 5.4 [2.46, 8.24] & 44.5 [36.84 , 51.93] & 16.3 [10.12, 21.72] & \textbf{0.75 [0.59 , 0.90]}f \\
    \bottomrule
    \end{tabular}
  \label{tab:bin_results}
  \vspace*{-5mm}
\end{table*}

Tables \ref{tab:ambi_results} and \ref{tab:bin_results} show the results obtained for the Ambisonic and binaural datasets, respectively. The results are reported for the testing set using jackknife estimation to obtain confidence intervals with significance level of 0.05.

As can be seen from most metrics, the performance of the models are significantly worse for the binaural dataset than for the Ambisonic dataset. For the multi-task approach, the error rate went up from the range of [0.63, 0.65] to [0.82, 0.87], whereas the $F_{1}$ score decrased from over 40\% to the range of [16.5, 20.0]\%. A similarly large effect can be seen for the localization metrics, where the DOA error ranges from 38.1° to 41.1° and the recall from 36.7\% to 45.6\%. This is a partly expected effect. Binaural recordings consist of only two channels as compared with four channels for FOA. Moreover, the cone of confusion effect and high directivity of the ears might impact the performance significantly. These problems can be largely overcome by utilizing a moving receiver as shown in \cite{Krause2024Moving, garcia2022binaural}. Interestingly, distance estimation has been affected to a much lesser extent than other tasks. For the MT approach, the distance error stays in the range between 0.98m to 1.04m, which is roughly a 10\% increase from the Ambisonic results. These results indicate that efficient distance estimation can be achieved even with binaural recordings, which shows a potential for future research.

As can be seen for both audio formats, for the multi-task approach most differences between loss functions appear for the distance error. This is expected given the ACCDOA branch uses the same loss for all scenarios. For MSE and MAE, the Ambisonic model achieves a similar distance error of 0.95~m and 0.93~m, respectively. This is a reasonable performance compared with other metrics. Introducing the distance scaling brings the error further down, to 0.89m for MSPE and 0.88m for MAPE.  Similarly, the binaural model achieves the lowest distance error of 0.98~m for MAPE. The SELD metrics show very limited influence of the distance loss on the training of the rest of the multi-task model. For FOA data, all scenarios achieve a fairly similar error rate between 0.63 and 0.65 and an $F_1$ score between 41.4\% for MSE and 44.1\% for MSPE. Interestingly, a higher impact can be seen for the localization metrics. Despite the DOA error staying between 21.6° and 23.2°, the recall varies between different scenarios. Compared with 61.0\% for MSE, the MAE loss results in a recall of 41.1\%, showing a decreased efficiency in estimated the correct number of sources. Also for binaural data, the MAE loss decreases the recall from 45.6\% to 36.7\%. However, for FOA the distance-scaled losses visibly improve the localization recall to 64.7\% and 64.5\% for MSPE and MAPE, respectively.

Larger differences between loss functions can be observed for the multi-ACCDDOA approach. The single task model trained with MSE achieved an ER and DOA error which are comparable with the multi-task approach. However, amongst all results obtained for the Ambisonic format, this model achieves the best $F_{1}$ score of 44.2\% and recall of 68.4\%. The distance error of 0.92m is similar to the one achieved for the multi-task with MSE. Similarly, the binaural model obtains the best $F_{1}$ score of 21.1\% and recall of 48.0\% using this method.

Unsurprisingly, changing the loss function to MAE affects all tasks to a much larger extent when using this method. First, the models achieves a distance error of 0.74m which is the best result across all experiments, showing that using the absolute error seems to be a better fit for estimating the distance. The loss affects positively the DOA error as well, for which the value goes down to 17.7°. However, detection metrics show a significant decrease of performance of the SED part - ER increases to 0.86, whereas the $F_{1}$ score goes down to 21.5\%. Similarly, recall goes down to 19.1\%, which is the worst result overall. For binaural data, the $F_{1}$ score drops to 5.4\%, which is an unacceptable performance. Such large differences between MSE and MAE show a discrepancy between the SDE and SELD parts, for which different loss functions are optimal.  

Across both datasets, best SELD performance is achieved for the multi-ACCDDOA approach using the MSE loss function. Regarding the fact that this method also allows for the detection of multiple overlapping of the same class, we consider this format superior to the multi-task approach. However, best distance error is achieved when using MAE loss. For future studies, we propose to investigate a mixed loss function, which would connect the benefits of using MSE for SELD and MAE for distance estimation. Alternatively, a different task definition might be used, combining the track-wise approach of multi-ACCDDOA with a multi-task output representation.

\section{Conclusions}
\label{conclusions}

In this study, we investigate the joint task of Sound Event Detection, Localization and Distance Estimation. We propose two methods of solving this problem - using a multi-task classwise approach and using the multi-ACCDDOA method. The methods are investigated using several loss functions for an Ambisonic and a binaural dataset. Our experiments show best results when using a multi-ACCDDOA approach with the MSE loss function. However, we note a disparity between SELD and distance estimation, where the latter performs better when using a MAE loss. Future studies could propose a mixed loss function to further improve the results.


\bibliographystyle{IEEEtran}
\small
\bibliography{conference_101719}

\begin{thebibliography}{10}
\providecommand{\url}[1]{#1}
\csname url@samestyle\endcsname
\providecommand{\newblock}{\relax}
\providecommand{\bibinfo}[2]{#2}
\providecommand{\BIBentrySTDinterwordspacing}{\spaceskip=0pt\relax}
\providecommand{\BIBentryALTinterwordstretchfactor}{4}
\providecommand{\BIBentryALTinterwordspacing}{\spaceskip=\fontdimen2\font plus
\BIBentryALTinterwordstretchfactor\fontdimen3\font minus \fontdimen4\font\relax}
\providecommand{\BIBforeignlanguage}[2]{{%
\expandafter\ifx\csname l@#1\endcsname\relax
\typeout{** WARNING: IEEEtran.bst: No hyphenation pattern has been}%
\typeout{** loaded for the language `#1'. Using the pattern for}%
\typeout{** the default language instead.}%
\else
\language=\csname l@#1\endcsname
\fi
#2}}
\providecommand{\BIBdecl}{\relax}
\BIBdecl

\bibitem{Virtanen2018Ananlysis}
T.~Virtanen, M.~D. Plumbley, and D.~Ellis, \emph{Computational Analysis of Sound Scenes and Events}.\hskip 1em plus 0.5em minus 0.4em\relax Springer, 2018.

\bibitem{speechrec}
M.~Woelfel and J.~McDonough, \emph{Distant speech recognition}.\hskip 1em plus 0.5em minus 0.4em\relax Wiley, 2009.

\bibitem{4058525}
J.~Hornstein, M.~Lopes, J.~Santos-Victor, and F.~Lacerda, ``Sound localization for humanoid robots - building audio-motor maps based on the hrtf,'' in \emph{2006 IEEE/RSJ International Conference on Intelligent Robots and Systems}, 2006, pp. 1170--1176.

\bibitem{opatka2011ApplicationOV}
K.~Łopatka, J.~Kotus, and A.~Czyżewski, ``Application of vector sensors to acoustic surveillance of a public interior space,'' \emph{Archives of Acoustics}, vol.~36, pp. 851--860, 2011.

\bibitem{5202720}
Y.-T. Peng, C.-Y. Lin, M.-T. Sun, and K.-C. Tsai, ``Healthcare audio event classification using hidden markov models and hierarchical hidden markov models,'' in \emph{2009 IEEE International Conference on Multimedia and Expo}, 2009, pp. 1218--1221.

\bibitem{Mesaros2018a}
A.~Mesaros, T.~Heittola, and T.~Virtanen, ``Acoustic scene classification: An overview of {DCASE} 2017 challenge entries,'' in \emph{16th International Workshop on Acoustic Signal Enhancement (IWAENC 2018)}, 2018.

\bibitem{Fonseca2018_DCASE}
E.~Fonseca, M.~Plakal, F.~Font, D.~P.~W. Ellis, X.~Favory, J.~Pons, and X.~Serra, ``General-purpose tagging of freesound audio with audioset labels: Task description, dataset, and baseline,'' in \emph{Proc. of the Detection and Classification of Acoustic Scenes and Events 2018 Workshop (DCASE2018)}, 2018, pp. 69--73.

\bibitem{krauseconv}
D.~Krause, A.~Politis, and K.~Kowalczyk, ``Comparison of convolution types in {CNN}-based feature extraction for sound source localization,'' in \emph{28th European Signal Processing Conference (EUSIPCO 2020)}, 2020, pp. 820--824.

\bibitem{Mesaros2019_TASLP}
A.~Mesaros, A.~Diment, B.~Elizalde, T.~Heittola, E.~Vincent, B.~Raj, and T.~Virtanen, ``Sound event detection in the {DCASE} 2017 {C}hallenge,'' \emph{IEEE/ACM Transactions on Audio, Speech, and Language Processing}, vol.~27, no.~6, pp. 992--1006, 2019.

\bibitem{Adavanne_2019}
S.~Adavanne, A.~Politis, J.~Nikunen, and T.~Virtanen, ``Sound event localization and detection of overlapping sources using convolutional recurrent neural networks,'' \emph{IEEE Journal of Selected Topics in Signal Processing}, vol.~13, p. 34–48, 2019.

\bibitem{9413473}
Y.~Cao, T.~Iqbal, Q.~Kong, F.~An, W.~Wang, and M.~D. Plumbley, ``An improved event-independent network for polyphonic sound event localization and detection,'' in \emph{ICASSP 2021 - 2021 IEEE International Conference on Acoustics, Speech and Signal Processing (ICASSP)}, 2021, pp. 885--889.

\bibitem{9053045}
T.~N. Tho~Nguyen, D.~L. Jones, and W.-S. Gan, ``A sequence matching network for polyphonic sound event localization and detection,'' in \emph{ICASSP 2020 - 2020 IEEE International Conference on Acoustics, Speech and Signal Processing (ICASSP)}, 2020, pp. 71--75.

\bibitem{shimada2021accdoa}
K.~Shimada, Y.~Koyama, N.~Takahashi, S.~Takahashi, and Y.~Mitsufuji, ``Accdoa: Activity-coupled cartesian direction of arrival representation for sound event localization and detection,'' in \emph{ICASSP 2021 - 2021 IEEE International Conference on Acoustics, Speech and Signal Processing (ICASSP)}, 2021, pp. 915--919.

\bibitem{9746384}
K.~Shimada, Y.~Koyama, S.~Takahashi, N.~Takahashi, E.~Tsunoo, and Y.~Mitsufuji, ``Multi-accdoa: Localizing and detecting overlapping sounds from the same class with auxiliary duplicating permutation invariant training,'' in \emph{ICASSP 2022 - 2022 IEEE International Conference on Acoustics, Speech and Signal Processing (ICASSP)}, 2022, pp. 316--320.

\bibitem{yiwere2019}
\BIBentryALTinterwordspacing
M.~Yiwere and E.~J. Rhee, ``Sound source distance estimation using deep learning: An image classification approach,'' \emph{Sensors}, vol.~20, no.~1, 2020. [Online]. Available: \url{https://www.mdpi.com/1424-8220/20/1/172}
\BIBentrySTDinterwordspacing

\bibitem{sobhdel2021}
\BIBentryALTinterwordspacing
A.~Sobhdel, R.~Razavi-Far, and S.~Shahrivari, ``Few-shot sound source distance estimation using relation networks,'' 2021. [Online]. Available: \url{https://arxiv.org/abs/2109.10561}
\BIBentrySTDinterwordspacing

\bibitem{kushwaha2023sound}
S.~S. Kushwaha, I.~R. Rom{\'{a}}n, M.~Fuentes, and J.~P. Bello, ``Sound source distance estimation in diverse and dynamic acoustic conditions,'' in \emph{{IEEE} Workshop on Applications of Signal Processing to Audio and Acoustics, {WASPAA} 2023}.\hskip 1em plus 0.5em minus 0.4em\relax {IEEE}, pp. 1--5.

\bibitem{Yiwere2017DistanceEA}
M.~Yiwere and E.~J. Rhee, ``Distance estimation and localization of sound sources in reverberant conditions using deep neural networks,'' in \emph{2017 International Journal of Applied Engineering Research}, 2017.

\bibitem{krause2021joint}
D.~A. Krause, A.~Politis, and A.~Mesaros, ``Joint direction and proximity classification of overlapping sound events from binaural audio,'' in \emph{2021 IEEE Workshop on Applications of Signal Processing to Audio and Acoustics (WASPAA)}.\hskip 1em plus 0.5em minus 0.4em\relax IEEE, 2021, pp. 331--335.

\bibitem{Krause2024Moving}
D.~A. Krause, G.~García-Barrios, A.~Politis, and A.~Mesaros, ``Binaural sound source distance estimation and localization for a moving listener,'' \emph{IEEE/ACM Transactions on Audio, Speech, and Language Processing}, vol.~32, pp. 996--1011, 2024.

\bibitem{yasuda2020sound}
M.~Yasuda, Y.~Koizumi, S.~Saito, H.~Uematsu, and K.~Imoto, ``Sound event localization based on sound intensity vector refined by dnn-based denoising and source separation,'' in \emph{2020 IEEE International Conference on Acoustics, Speech and Signal Processing (ICASSP)}, 2020, pp. 651--655.

\bibitem{garcia2022binaural}
G.~Garc{\'\i}a-Barrios, D.~A. Krause, A.~Politis, A.~Mesaros, J.~M. Guti{\'e}rrez-Arriola, and R.~Fraile, ``Binaural source localization using deep learning and head rotation information,'' in \emph{2022 30th European Signal Processing Conference (EUSIPCO)}.\hskip 1em plus 0.5em minus 0.4em\relax IEEE, 2022, pp. 36--40.

\bibitem{paszke2019pytorch}
\BIBentryALTinterwordspacing
A.~Paszke, S.~Gross, F.~Massa, A.~Lerer, J.~Bradbury, G.~Chanan, T.~Killeen, Z.~Lin, N.~Gimelshein, L.~Antiga, A.~Desmaison, A.~Köpf, E.~Yang, Z.~DeVito, M.~Raison, A.~Tejani, S.~Chilamkurthy, B.~Steiner, L.~Fang, J.~Bai, and S.~Chintala, ``Pytorch: An imperative style, high-performance deep learning library,'' 2019. [Online]. Available: \url{https://doi.org/10.48550/arXiv.1912.01703}
\BIBentrySTDinterwordspacing

\bibitem{Shimada2023starss23_arxiv}
\BIBentryALTinterwordspacing
K.~Shimada, A.~Politis, P.~Sudarsanam, D.~Krause, K.~Uchida, S.~Adavanne, A.~Hakala, Y.~Koyama, N.~Takahashi, S.~Takahashi, T.~Virtanen, and Y.~Mitsufuji, ``{STARSS23}: An audio-visual dataset of spatial recordings of real scenes with spatiotemporal annotations of sound events,'' \emph{In arXiv e-prints: 2306.09126}, 2023. [Online]. Available: \url{https://arxiv.org/abs/2306.09126}
\BIBentrySTDinterwordspacing

\bibitem{politis2021dataset}
A.~Politis, S.~Adavanne, D.~Krause, A.~Deleforge, P.~Srivastava, and T.~Virtanen, ``A dataset of dynamic reverberant sound scenes with directional interferers for sound event localization and detection,'' in \emph{Proceedings of the 6th Detection and Classification of Acoustic Scenes and Events 2021 Workshop (DCASE2021)}, pp. 125--129.

\bibitem{fonseca2022fsd50k}
E.~Fonseca, X.~Favory, J.~Pons, F.~Font, and X.~Serra, ``{FSD50K}: an open dataset of human-labeled sound events,'' \emph{IEEE/ACM Transactions on Audio, Speech, and Language Processing}, vol.~30, pp. 829--852, 2022.

\bibitem{spaudiopy}
\BIBentryALTinterwordspacing
C.~Hold, ``{spaudiopy},'' 2023. [Online]. Available: \url{https://github.com/chris-hld/spaudiopy/tree/master}
\BIBentrySTDinterwordspacing

\bibitem{ambisonics}
F.~Zotter and M.~Frank, \emph{Ambisonics}.\hskip 1em plus 0.5em minus 0.4em\relax Springer Cham, 2019.

\bibitem{politis2020overview}
A.~Politis, A.~Mesaros, S.~Adavanne, T.~Heittola, and T.~Virtanen, ``Overview and evaluation of sound event localization and detection in dcase 2019,'' \emph{IEEE/ACM Transactions on Audio, Speech, and Language Processing}, pp. 1--14, 2020.

\end{thebibliography}

\end{document}